\documentclass[preprint2]{aastex631}
\usepackage{graphicx}	
\usepackage{amsmath}	

\begin{document}

\title{Bayesian Insights into post-Glitch Dynamics: Model comparison and parameter constraint from decades long observation data of the Crab pulsar}

\author[0000-0001-6406-1003]{Chun Huang}
\affiliation{Physics Department, Central China Normal University, Luoyu Road, 430030, Wuhan, China}
\affiliation{Physics Department and McDonnell Center for the Space Sciences, Washington University in St. Louis; MO, 63130, USA}
\author[0000-0001-8868-4619 ]{Xiao-ping Zheng}
\affiliation{Physics Department, Central China Normal University, Luoyu Road, 430030, Wuhan, China}
\email{zhxp@ccnu.edu.cn (XPZ)}



\begin{abstract}
The Crab Pulsar has exhibited numerous glitches accompanied by persistent shifts in its spin-down rate. The explanation of the observed persistent shifts remain a challenge. We perform a detailed Bayesian analysis to compare four data-fitting models,  ranging from a simple linear model to more complex power-law and logarithmic models, using a dataset of observed glitches and persistent shifts. Our results show the large observed events are difficult to explain by the usually assumed linear model due to starquakes. A particularly notable finding is that the logarithmic model provides the best fit to the observation data but the two power-law models show  a close tie to it. Detail differences of these models may be further clarified  by the understanding of internal physics of neutron stars.
\end{abstract}



\section{Introduction} \label{sec:intro}
The Crab Pulsar has been observed for over half a century, particularly through radio band observations( eg. \cite{Lyne1992,lyne2011,radio2017,radio2024}), resulting in a substantial accumulation of data on its rotational evolution, with significant contributions from the Jordrell Bank Observatory \cite{2011MNRAS.414.1679E,2015MNRAS.446..857L,2022MNRAS.510.4049B}. Nearly 30 instances of a sudden increase in the pulsar's spin frequency, known as glitches, have been recorded. These glitches are typically accompanied by a permanent shift in the spin-down rate of the pulsar(see \cite{1985ApJ...288..191A,1996ApJ...459..706A,2017MNRAS.469.4183A,2019arXiv190612060W,2020ApJ...896...55G}). Despite extensive observations, the underlying mechanisms driving these glitches remain unclear. One hypothesis suggests that glitches may result from starquakes \cite{Baym1971,1972NPhS..237...83P,1994MNRAS.269..849A,Link1998,2002MNRAS.333..613L,2015MNRAS.446..865K,2018MNRAS.473..621A,2023MNRAS.520.4289L,2020MNRAS.493L..98S} on the neutron star, while another model attributes them to the superfluidity of the neutron star's matte \cite{1975Natur.256...25A,1976Ruderman,1977ApJ...213..527A,1981IAUS...95..299A,1984ApJ...276..325A,1985alpar,1993ApJ...403..285L,2002MNRAS.336..211L,2009PhRvL.102n1101G,2014ApJ...789..141L,2016PhRvL.117w2701W}, which may also account for the abrupt increase in spin frequency.

The Crab Pulsar is particularly active in exhibiting glitches, likely due to its relative youth. Detailed modeling of glitch data from this pulsar is therefore crucial for understanding the phenomenon of pulsar glitches \cite{1985alpar,1996ApJ...459..706A,2000MNRAS.315..534L,2003ApJ...595.1052C,2019arXiv190612060W,2020MNRAS.493L..98S}. In this study, we focus primarily on the  persistent post-glitch shifts in spin-down rate following glitches, with a particular emphasis on the relationship between these  effects and glitch sizes. This relationship may provide a useful probe for distinguishing among various glitch models, as the recovery behavior following a glitch can be modeled in greater detail due to its extended timescale, thereby offering rich information about the glitch mechanisms.

To analyze the glitch data, we will employ modern Bayesian techniques, which are well-suited for extracting information from observational datasets and for conducting model comparisons \cite{kass}. Bayesian inference allows for the systematic comparison of different  models on a consistent basis, and also facilitates the constraints of model parameters within various frameworks. The objective of this study is to bridge the gap between  models and observational data through systematic statistical modeling of post-glitch activity in the Crab Pulsar.

This paper is organized as follows: In Section 2, we will discuss the datasets used and the Bayesian framework implemented, along with the modeling motivations for the different data-fitting models. Section 3 will present the posterior results of the Bayesian inference, comparing the Bayesian evidence across different models, and explore the inter-model numerical conversion relations, to emphasis the deep connection between each two models. In section 4, the discussion of recent observation from 2017 event will be presented
\begin{table*}
\centering
\setlength{\tabcolsep}{7mm}{\begin{tabular}{l c} 
\hline\hline
\text { Glitch size $\Delta\nu$ $(\mu \mathrm{Hz})$} & {Persistent shift $\Delta \dot{\nu}_p$ $\left(10^{-15} \mathrm{~s}^{-2}\right)$} \\
\hline

 1.08(1) & -112(2) \\
2.43(1) & -150(5)\\ 
1.133(3) & -116(5)\\
0.20(1) & -25(3)\\
0.855(3) & -53(3)\\
0.94(3)& -70(10)\\
0.151(3)&-8(2)\\
6.595(20) & 250(20)\\
0.65(1)& - 30(5)\\
1.46(1)& -132(5)\\
\hline\hline
\end{tabular}}
\caption{Summary for the data-set of observed glitch size and the persistent shift. See Table 3 in \cite{2015MNRAS.446..857L}}
\label{tab1:dataset}
\end{table*}
\section{Method description} \label{sec:style}
The Bayesian technique requires several ingredients: observation data-set, the theoretical models, prior setting of all the model parameters,  likelihood set-up and nested sampling algorithm. With accumulative observations of several decades, the observation data of the Crab pulsar already very significant, In \cite{2015MNRAS.446..857L}, people summarized the 45 years of glitch observation on the Crab pulsar, with detailed record of the glitch sizes and the persistent shifts of  post-glitch spin-down rates. (See Table 3 of \cite{2015MNRAS.446..857L}). Our primary focus is the relation between persistent shift $\Delta \dot{\nu}_p$ and glitch size $\Delta \nu$. The data is listed in the table \ref{tab1:dataset}. Larger glitch size normally corresponds to a larger step, this general correlation, in \cite{2015MNRAS.446..857L} was fitted by a linear model. While the detail treatment of this correlation is crucial for distinguishing different  models, and only the linear model will prohibit a very bad explanation of the observation data. To explore this correlation systematically by modern statistical approach motivates this study. 
\subsection{Fitting models}\label{models}
In this current study, four different models will be  examined by the observation data, regardless of realistic physics. (a) The linear fitting model $\Delta \dot{\nu}_p = k_{0} \Delta\nu + b_{0}$ (denoted as M1),  where $k_0$ and $b_0$ are free parameters that need to explored by Bayesian technique. This model was presented by \cite{2015MNRAS.446..857L}.(b) The power-law fitting model $\Delta \dot{\nu}_p = k_{1}\Delta \nu^{\gamma_{1}}$ (denoted as M2), where $k_1$ is the proportional constant, $\gamma_1$ is the power-law index, these are the free parameters in this model. This model release the polynomial index of linear model from 1 to a free parameter to embed possible nonlinear correlation between glitch size and persistent shift. (c) The modified power-law fitting model $\Delta \dot{\nu}_p = \Delta\nu^{\gamma_{1}}(\alpha_1 + \xi_1 \Delta\nu^{\gamma_2})^{-1}$(denoted as M3),
where the modified term on the simple power law $\xi_1 \Delta\nu^{\gamma_2}$ is a small quantity compared to the coefficient $\alpha_1$. Here there are four free parameters $\gamma_1$, $\gamma_2$, $\alpha_1$ and $\xi_1$. This model naturally seems to include some  higher-order effects of possible physical process. (4) The Log fitting model, $\Delta \dot{\nu}_p = k_{2}\ln(\alpha_2\Delta \nu + \xi_2)$ (denoted as M4) , where the $k_2$, $\alpha_2$ and $\xi_2$ are the free parameters. This model is currently lack of appropriate motivation, however, based on the observation of the overall trend of the data-set, this is a reasonable pick mathematically. In the following study, Bayesian analysis is performed to constrain the model parameters and the inter-connection among these models will be discussed in detail. 
\subsection{Bayesian inference set-up}
Prior setting of these models reflects our preknowledge about the fitting free parameters, this is the essential components of Bayesian analysis. For linear model, the prior for $k_0$ and $b_0$ are uniform probablity distribution from 0 to 300, and -300 to 300, which denoted as $\mathcal{U}(0,300)$ and $\mathcal{U}(-300,300)$. The power-law model priors defined as the prior of $k_1$ is $\mathcal{U}(0,300)$, for power-law index $\gamma_1$ is $\mathcal{U}(0,3)$. In modified power-law model, we set $\mathcal{U}(0,3)$ for $\gamma_1$ and $\gamma_2$, the $\alpha_2$ and $\xi_2$ prior are both $\mathcal{U}(0,0.1)$.
\begin{figure}
	\centering
	\includegraphics[scale=0.50]{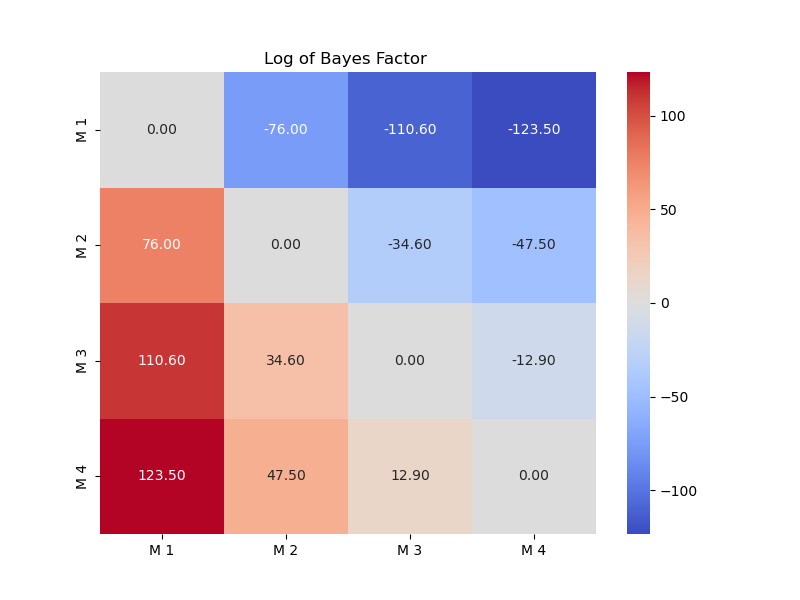}
	\caption{The heatmap of different Bayes factors given different models: (1) M1: Linear model. (2) M2: Power-law model. (3) M3: Modified power-law model. (4) M3: Log model}
	\label{Bayes_fac}
\end{figure} 
Regarding the likelihood set-up, our likelihood should reflect the observation data and their uncertainties, and as the constraint condition for reshaping the prior space of each models. Here, each individual observation of the glitch event with their accompanied persistent shift will associate with a likelihood function then compose all of them to a single global likelihood. As the uncertainty of glitch size measurement and persistent step width given, we view each individual observation as a Gaussian likelihood function with their 1-sigma width is determined by the uncertainty number. Every likelihood computation will be carried as following: Sampling the glitch size of the correlated persistent step observation via the given uncertainty-defined Gaussian distribution function. Then take the sampled glitch size into our prior-defined fitting model get a corresponded persistent shift value. Put these values inside of the Gaussian likelihood defined by Persistent shift observation, compute the likelihood from this observation then draw the posterior from this process by multiplying the likelihood with prior which defined the theoretical model utilized in this computation. This process could be expressed as a formula:
\begin{equation}
\begin{aligned}
& p(\boldsymbol{\theta} \mid \boldsymbol{d}, \mathcal{M}) \propto p(\boldsymbol{\theta} \mid \mathcal{M})  \\
& \times \prod_j p\left( \Delta\dot{\nu}_{p,j} \mid d_{\mathrm{radio}, \mathrm{j}}\right)p\left( \Delta\nu_{j} \mid d_{\mathrm{radio}, \mathrm{j}}\right)
\end{aligned}
\end{equation}
The meaning of this euqation is, for given model and data-set the posterior probability distribution of the model parameters are defined, according to Bayes' theorem, as the product of the prior distribution of these parameters for a given model and the prior probability distribution of glitch size $\Delta\nu$ and persistent shift $\Delta\dot{\nu}_{p}$ giving the radio observation data set $d_{\mathrm{radio}}$ as the likelihood.

All analysis in this study were performed using the \textit{CompactObject} package, created by the authors. This open-source tool is designed for applying Bayesian constraints to the interiors of neutron stars, based on observational data. This research marks the first time the package has been applied to another physics problem other than equation of state inference. Previous studies using this package include those referenced in \cite{Huang:2023grj}. The sampling process employed in this research utilizes \textit{UltraNest}, which uses a highly efficient step-sampler for nested sampling \cite{2021JOSS....6.3001B}. We configured the model comparisons with 50,000 live points for each model, a setting that dictates the extent of Bayesian exploration and guarantees a fair comparison among models. The average sampling duration for these models on a standard personal laptop is about 600 seconds, demonstrating the computational viability of this method given the manageable size of the observational data.
\begin{table}
\centering
\setlength{\tabcolsep}{7mm}{\begin{tabular}{l c} 
\hline\hline
\text { Models} & { Baysian evidence }($\log (Z)$) \\
\hline

 M1 & -280.4 $\pm$ 0.032\\
M2 & -204.4 $\pm$ 0.023\\ 
M3 & -167.7 $\pm$ 0.046\\
M4 & -156.9 $\pm$ 0.018\\
\hline\hline
\end{tabular}}
\caption{Summary for the Bayesian evidence for different models: (1) M1: Linear model. (2) M2: Power-law model. (3) M3: Modified power-law model. (4) M4: Log model }
\label{tab2:evidence}
\end{table}
\begin{table*}
\centering
\setlength{\tabcolsep}{7mm}{\begin{tabular}{l c} 
\hline\hline
\text { Models} & { Parameters posteriors }($\log (Z)$) \\
\hline

 M1 & $k_0 = 68.94^{+1.61}_{-1.61}$, $b_0 = 11.59^{+1.60}_{-1.60}$ \\
M2 & $k_1 = 87.31^{+1.23}_{-1.25}$, $\gamma_1 = 0.74^{+0.02}_{-0.02}$\\ 
M3 & $\gamma_1 = 0.44^{+0.04}_{-0.05}$, $\gamma_2 = -0.096^{+0.032}_{-0.044}$,
$\xi_1 =0.0035^{+0.0004}_{-0.0006}$, $\alpha_1 =0.0044^{+0.0010}_{-0.0010}$\\
M4 & $k_2 = 120.59^{+13.80}_{-12.21}$, $\alpha_2 = 0.89^{+0.03}_{-0.03}$,
$\xi_2 =1.26^{+0.23}_{-0.20}$\\
\hline\hline
\end{tabular}}
\caption{Summary for the Bayesian evidence for different models: (1) M1: Linear model. (2) M2: Power-law model. (3) M3: Modified power-law model. (4) M3: Log model }
\label{tab3:posterior}
\end{table*}
\begin{figure}
	\centering
	\includegraphics[scale=0.50]{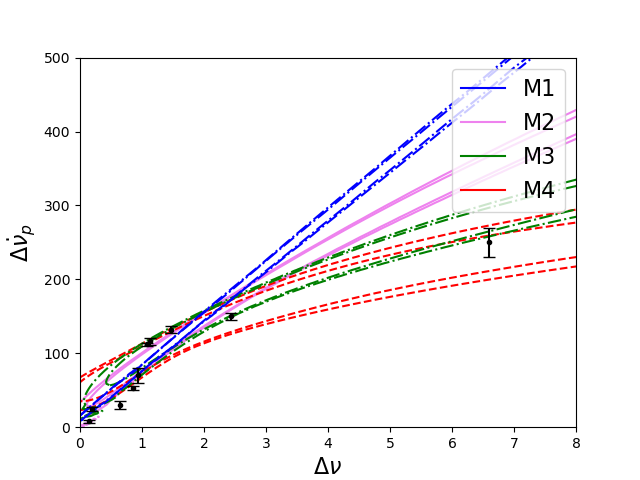}
	\caption{The projection of posterior samples in $\Delta \nu - \Delta \dot{\nu}_p$ given different models: (1) M1: Linear model. (2) M2: Power-law model. (3) M3: Modified power-law model. (4) M3: Log model. Where$\Delta \nu$ is in $\mu \mathrm{Hz}$, $ \Delta \dot{\nu}_p$ is in $10^{-15} \mathrm{~s}^{-2}$. The contour levels are 68\% and 84\% separately from inner to outer.}
	\label{projection}
\end{figure} 
\begin{figure}
	\centering
	\includegraphics[scale=0.50]{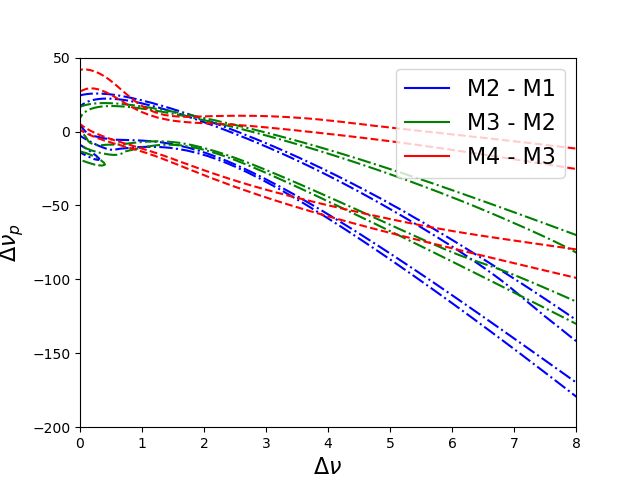}
	\caption{The projection of residual of each pair of models in $\Delta \nu - \Delta \dot{\nu}_p$ given different models pair: (1) M1 and M2 (blue) (2) M2 and M3 (green)(3) M3 and M4 (red). Where$\Delta \nu$ is in $\mu \mathrm{Hz}$, $ \Delta \dot{\nu}_p$ is in $10^{-15} \mathrm{~s}^{-2}$. The contour levels are 68\% and 84\% separately from inner to outer.}
	\label{correction}
\end{figure} 

\section{Inference result}
Different model choices result in varying Bayesian evidence, reflecting how well each model explains the data. By analyzing this Bayesian evidence, we can perform model comparisons and draw data-driven conclusions about which models are favored by the observational data. Table \ref{tab2:evidence} lists the logarithms of the Bayesian evidence for each models. 

Model M1,  the linear model that  probably matches the starquake hypothesis\cite{}, is the least favored by the observation data, with a very low value of -280.4. In contrast, the non-linear model, M2, shows a significant improvement in Bayesian evidence, rooted in the fact that the power-law function fits the large glitch-accompanied persistent shift better. Introducing a higher-order correction to Model M2, that is Model M3, further improves the Bayesian evidence significantly. 

The final model, M4,  particularly noteworthy because its functional form is quite another, yet it demonstrates the highest explanatory power for the observational data. This logarithmic-like dependence might represent some missed correction to the power-law models or a different mechanism altogether. In the following section, we will statistically show the convergence of each model, aiding in understanding the potential origin of this purely mathematical model.

However, it is important to note that, across all four models, the Bayesian evidence remains quite low, with the maximum value being -156.9. Considering this is the $\log(Z)$, the exponential of this value is extremely small. This may arise from the fact that the observational data for small glitch sizes accompanied by persistent shifts are well-observed, resulting in very small uncertainties, yet these phenomena seem challenging to explain smoothly using any elementary functional form. This raises questions about the reliability of the observations or suggests that the small glitch-related persistent shifts may require a more sophisticated model for adequate explanation.

In Figure \ref{Bayes_fac}, the heatmap of Bayes' factors for different model pairs is presented. The Bayes factor is a standard method for model selection (see \cite{kass}). A Bayes factor greater than 1 but less than 3.2 for model A versus model B suggests that the observations moderately favor model A, though this is not definitive. If the Bayes factor exceeds 3.2, it provides strong evidence that the data prefer model A. In this heatmap, the natural logarithm of the Bayes' factors is shown.

This map indicates that from M1 to M4, each successive model represents a significant improvement over the previous one. The conclusion is robust, as the Bayes' factors are exceptionally large, demonstrating that the measurements are sufficiently precise to distinguish among different theoretical models in this context. This provides overwhelmingly strong evidence from Bayesian inference to rule out the validity of the linear model. The large Bayes' factor suggests that the linear model fails to capture some dominant feature of the system.

When comparing M3 to M2, the large Bayes factor indicates that the higher-order correction to an idealized power-law model is crucial for explaining the data. Even though this correction is theoretically minor, the data are already significant enough to distinguish this model from the one without the correction. The improvement from M3 to M4 is relatively smaller but still extremely strong. This gives us additional motivation to explore the physical meaning behind this model, as the improvement is critical, even though the model is currently only mathematical in nature.

The detailed posterior contour plots are provided in the supplementary materials. While the parameter space itself may be less interesting in this study—since the parameters lack direct physical interpretation and are purely model parameters—they become crucial in future theoretical discussions. However, translating these posterior samples back to the $\Delta \nu - \Delta \dot{\nu}_p$ space is of interest, as it allows us to see the distinct regions in which these models reside. By comparing these regions with the original observational data, we can gain insights into the performance of the various models and correlate these findings with the Bayesian evidence discussed earlier.

All the parameters posteriors are summarized in Table \ref{tab3:posterior}. In Figure \ref{projection}, we illustrate the projection of these posterior distributions of the four different models onto the $\Delta \nu - \Delta \dot{\nu}_p$ plane. Our analysis shows that for all theoretical models, when the glitch size is smaller than approximately 3 $\mu$Hz, their performance is comparable, as they tend to occupy similar regions. However, for larger glitches, particularly the last data point with a glitch size exceeding 6 $\mu$Hz, the observed persistent shift is much smaller than predicted by M1 and M2, with the observation falling outside their 84\% credible interval.

Comparing the overall bandwidth of different models, we observe that the 84\% credible interval bandwidth increases sequentially from M1 to M4. This reflects that, from M1 to M4, the models expand the available parameter space to better explain the observational data. This expansion is also reflected in the Bayesian evidence, which measures the volume of the posterior parameter space.

An interesting observation is the sequentially increasing trend in Bayesian evidence from M1 to M4, which aligns with the models' progressively better performance in explaining the last large glitch observation. This suggests that the large glitch observation is the key factor in distinguishing among the models from a statistical perspective. Notably, this large data point cannot be well-reproduced without compromising the fit for the small glitch accompanied by persistent shift observations, highlighting the challenge in balancing model accuracy across different scales of glitch sizes.

This increasing trend of approaching the last observation also suggests an inter-model numerical convergence relationship. Each model performs similarly at lower glitch sizes, but in the larger glitch region, from M1 to M4, the models sequentially approach the largest data point. In Figure \ref{correction}, we show the posterior projection residuals between each pair of models, defined as $\Delta \dot{\nu}_{p,M_{i}} - \Delta \dot{\nu}_{p,M_{j}}$, where the i-th model's predicted persistent shift is subtracted from the j-th model's prediction. This allows us to clearly see that all model comparisons approach zero for smaller glitches. However, for larger glitch sizes, the correction from the i-th model to the j-th model predicts a lower $\Delta \dot{\nu}_p$.

Statistically, this indicates that the models with better performance and larger Bayesian evidence account for higher-order corrections that become significant only in the case of large glitches. For the M3-M2 pair, this higher-order correction is theoretically well understood—it stems from new physics beyond the power-law process. As we show, this correction is relatively small for small glitches but becomes dominant for large glitch sizes.

For the M4-M3 pair, this higher-order correction is still evident. This suggests that the reason M4 outperforms the other models in Bayesian inference is likely due to this higher-order correction, assuming both models share the same foundational framework. However, the physical meaning behind the higher-order correction that M4 accounts for remains unclear and will require further modeling work. Despite this, the statistical behavior of the correction is clear from our analysis.
\begin{table}
\centering
\setlength{\tabcolsep}{7mm}{\begin{tabular}{l c} 
\hline\hline
\text { Models} & { Baysian evidence }($\log (Z)$) \\
\hline

 M1 & -635.1 $\pm$ 0.026\\
M2 & -250.8 $\pm$ 0.028\\ 
M3 & -173.3 $\pm$ 0.052\\
M4 & -173.0 $\pm$ 0.089\\
\hline\hline
\end{tabular}}
\caption{Summary for the Bayesian evidence for different models after implementing the 2017 glitch event observation: (1) M1: Linear model. (2) M2: Power-law model. (3) M3: Modified power-law model. (4) M3: Log model }
\label{tab3:evidence_new}
\end{table}
\begin{figure}
	\centering
	\includegraphics[scale=0.50]{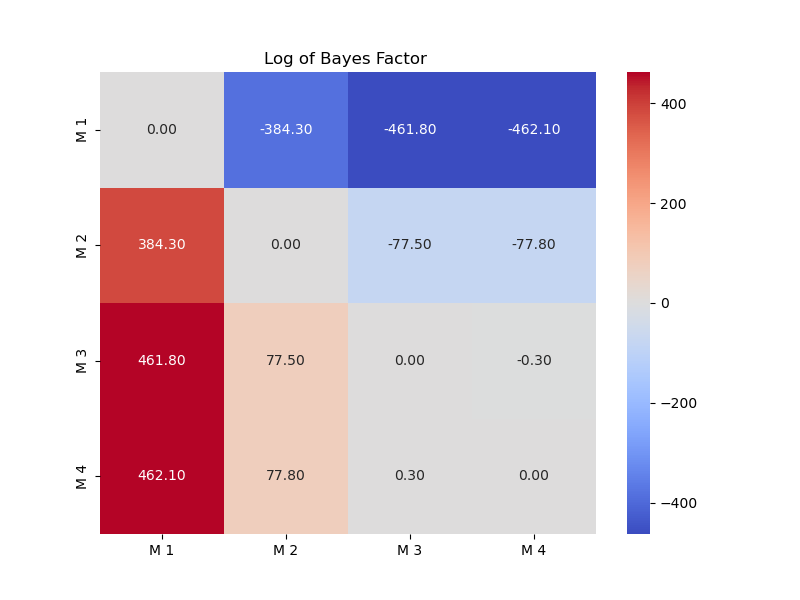}
	\caption{The heatmap of different Bayes factors given different models after implementing the 2017 glitch event observation: (1) M1: Linear model. (2) M2: Power-law model. (3) M3: Modified power-law model. (4) M3: Log model}
	\label{Bayes_fac_new}
\end{figure} 
\begin{figure*}
	\centering
	\includegraphics[scale=0.50]{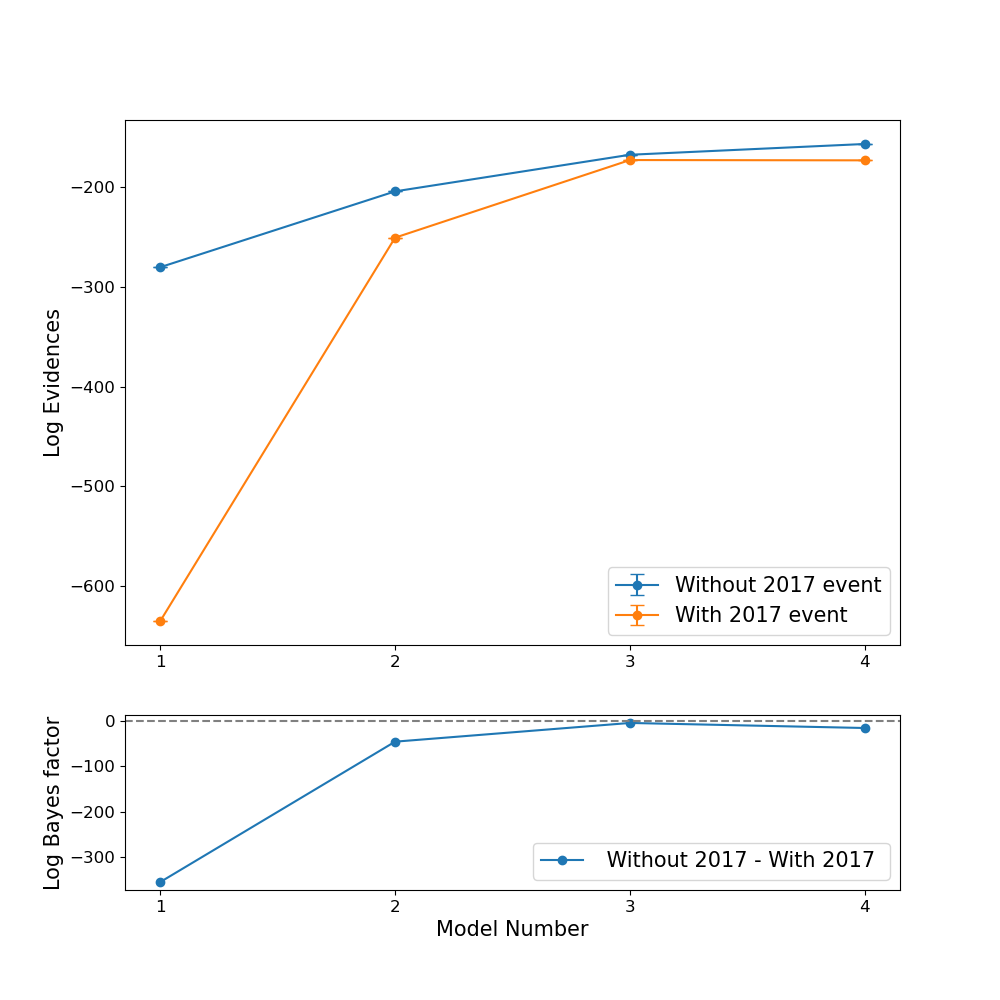}
	\caption{The distribution of the Bayesian evidence accross different models, comparing the result with and without 2017 observation for (1) M1: Linear model. (2) M2: Power-law model. (3) M3: Modified power-law model. (4) M3: Log model}
	\label{compare}
\end{figure*} 
\begin{figure*}
\centering
{\includegraphics[width=0.45\textwidth]{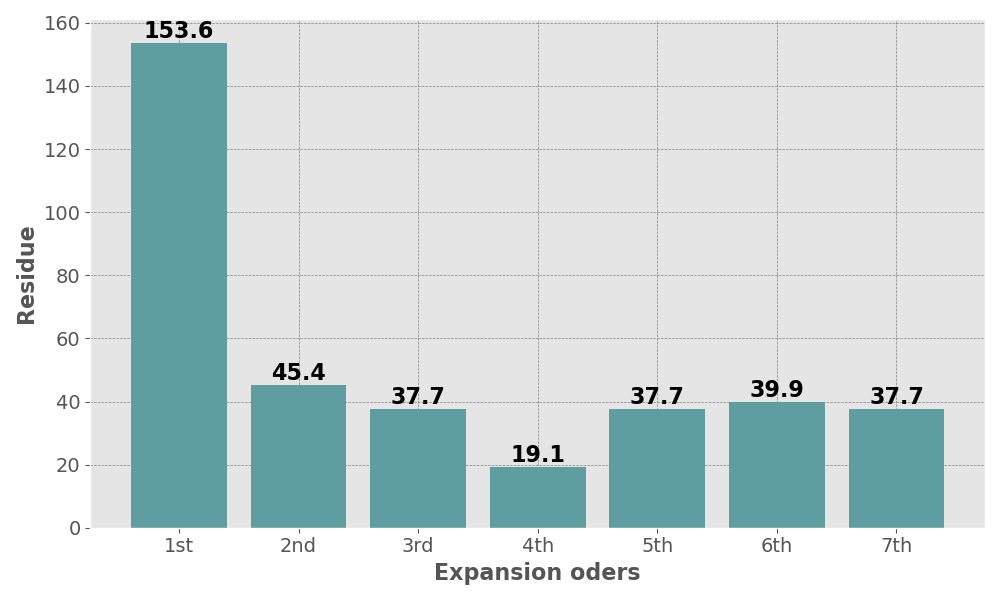}}
{\includegraphics[width=0.45\textwidth]{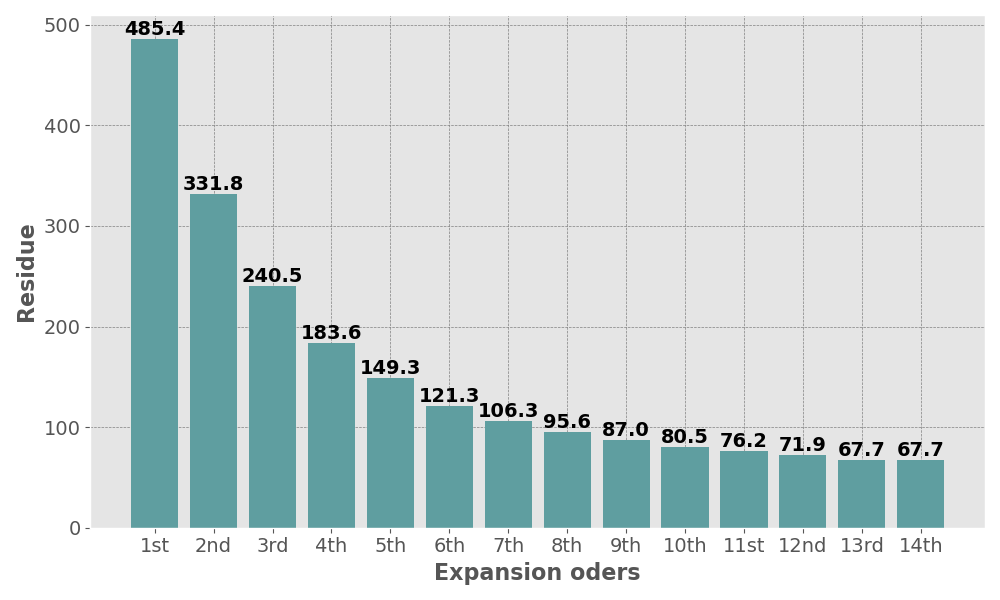}}
\caption{The Residue comparison (1) left panel: between the M3 and M2, (2) right panel: between the M4 and M3, The unite of the residue is $10^{-15}$$\mu$Hz$\cdot$$s^{-2}$}
\label{fig:residue_comparison}
\end{figure*}
\section{Test in ultra glitch}
In 2017, a new glitch event was recorded, accompanied by a persistent shift in the spin-down rate of the star. This glitch was obscured by the occurrence of three much smaller glitches in later about 600 days \cite{2018MNRAS.478.3832S,2020MNRAS.491.3182B,2022MNRAS.510.4049B,2021MNRAS.505L...6S} and hence they were incorported into an effective event as  16.43 $\mu$Hz,  with a persistent shift of 434(8) in units of $\left(10^{-15} \mathrm{~s}^{-2}\right)$\cite{}. Since this event is more than twice the glitch size of the previous largest event, it presents a significant test of the current  models discussed in this study. As a result, a detailed Bayesian inference was conducted with this new observation added to the dataset.

Table \ref{tab3:evidence_new} summarizes the updated Bayesian evidence for the four  models  inclusion of the 2017 event. The overall sequence of Bayesian evidence remains consistent: from M1 to M4, the Bayesian evidence improves, supporting the conclusion that higher-order corrections are incorporated into the models, making them more accurate. Notably, the Bayesian evidence for the linear model has dropped to an extremely small value, which allows us to confidently exclude this linear model as an explanation and provides strong motivation to propose new theoretical models to interpret this observation.

In Figure \ref{Bayes_fac_new}, we present an updated heatmap of the Bayes' factors. All the Bayes' factors in the lower-left corner have increased, reflecting the fact that the inclusion of this new data makes the distinction among models even stronger. The only exception is the M4/M3 comparison, where the logarithm of the Bayes' factor decreases from 12.90 to 0.3 after incorporating the 2017 event. This is a particularly interesting observation, as it suggests that the 2017 event might be better understood by M3. This single observation also is significant enough to support the overall conclusion. The dataset as a whole still  favors Model 4, as the Bayes' factor remains larger than 1, providing potential evidence that M4 is still the best model to explain the entire dataset.

In Figure \ref{compare}, we compare the Bayesian evidence of different models with and without the 2017 event included. The conclusion is that, with the inclusion of this new observation, all models more and less face challenges in explaining this large event, as the logarithms of Bayesian evidence have dropped, and the posterior space has become more constrained. Especially, M1 and M2 exhibit a significant drop in evidence, suggesting that these two models are strongly constrained and potentially excluded by the new observation.

For M3 and M4, the decrease in Bayesian evidence is not as pronounced, particularly when compared to M1 and M2. If we compute the Bayes' factor for the same model with and without the 2017 event, the logarithm of the Bayes' factor is relatively small. Interestingly, the Bayes' factor for M4 is even smaller than for M3, consistent with our earlier analysis showing that only the M4/M3 Bayes' factor decreased when comparing the scenarios with and without the 2017 event.
\section{Inter-models numerical conversions}
Most of the models considered in this study are primarily mathematical in nature, with the physical interpretation of their parameters remaining largely undefined. However, even when treating these models as purely mathematical constructs, we can still gain insights into their characteristics by performing numerical expansions to various orders. This approach allows us to explore the inter-model conversions based on these expansions. It is possible that truncating higher-order effects could reduce the more data-favored models to resemble the less-favored ones.

In this analysis, we focus exclusively on the best-fitting parameter sets for the different models, specifically selecting the optimal parameter combinations listed in Table \ref{tab3:posterior}. The methodology is as follows: for each model pair, we subtract the less-favored model from the more data-favored model, and then numerically integrate this difference from zero to the final glitch observation point. This integration provides a global residue that quantifies the discrepancy between the models. Our primary focus is on the M3-M2 and M4-M3 model pair conversions. The M2-M1 pair is straightforward, as the first-order expansion of M2 can easily revert to the linear M1 model.

For the M3-M2 pair, see the left panel of Figure \ref{fig:residue_comparison}. The x-axis represents the expansion orders of M3, where the expansion is based on a Taylor series. We observe that the residue is minimized at the fourth-order expansion; beyond this point, increasing the expansion order causes the residue to rise again. This suggests that the fourth-order expansion of M3 is the best numerical fit for M2. The superior performance of M3 can be attributed to its incorporation of higher-order modifications from unknown physical sources, thereby numerically demonstrating the correlation between different models.

The residue for the M4-M3 pair is shown in the right panel of Figure \ref{fig:residue_comparison}. This comparison is more complex because the usual Taylor expansion of the M4 logarithmic model does not monotonically increase with additional terms—each positive term is followed by a negative one, making the expansion less effective at reproducing the less-favored model. To address this, we employ a special expansion of the logarithmic function:
\begin{equation}
k_2 \ln \left(\xi_2+\alpha_2 \Delta \nu\right) = k_2 \sum_{m=0}^{\infty} \frac{\left(\left(\xi_2+\alpha_2 \Delta \nu\right)-1\right)^m}{m\left(\xi_2+\alpha_2 \Delta \nu\right)^m}
\end{equation}
This expansion holds true as long as the condition $\operatorname{Re}\left(\xi_2+\alpha_2 \Delta \nu\right)>1 / 2$ is satisfied. This condition is automatically met across the entire $\Delta \nu$ range, given that $\xi_2$ is set to 1.26 and the minimum value of $\Delta \nu$ is zero. Using this expansion, and as shown in Figure \ref{fig:residue_comparison}, the residue stabilizes after 14 terms. Thus, we can assert that the M3 modified power-law model is equivalent to the 14th-order expansion of the M4 logarithmic model. The better performance of M4 on explaining the observation data can be attributed to its inclusion of higher-order corrections absent in the previous model.

This finding is particularly intriguing because, even without an explicit theoretical framework to derive the logarithmic model, we can numerically establish the connection between existing models and identify the orders of the source of the corrections. Interpreting the physical significance of these higher-order terms will require further theoretical modeling.

\section{Conclusion}
In this study, we conducted a detailed Bayesian analysis of the glitch-induced persistent increases in the spin-down rate of the Crab pulsar. By systematically comparing four models, ranging from a simple linear model (M1)  to more sophisticated power-law and modified power-law models (M2 and M3), as well as a logarithmic model (M4). We hope to understand how well these models explain the observational data.

Our results indicate that higher-order corrections are crucial for accurately explaining the persistent shift accompanying larger glitches. From M1 to M4, the Bayesian evidence increases, showing that the models progressively improve in capturing the post-glitch behavior, especially when considering larger glitches. The linear model (M1) is significantly disfavored by the data. This highlights the need for more complex models that incorporate effects like superfluidity and its associated corrections, as represented by M2 and M3.

 The recent 2017 glitch is a extremely large event. Its inclusion in our analysis further constrained the models, particularly M1 and M2, which were strongly disfavored. However, M3 and M4 remain competitive, with M4 still slightly outperforming M3 statistically and  M3  maybe better capturing the nuances of this 2017 largest glitch event property.  Further modeling work is needed to fully understand success of model M4.

The post-glitch behavior of the Vela is a classical relaxation process, associated with a response to vortex creep. Unlike this standard case, the Crab pulsar cannot relax toward the pre-glitch state after a glitch and has a persistent torque increase. The effect was generally attributed to a external torque variation triggered by starquakes \cite{1992ApJ...390L..21L, 1997ApJ...478L..91L, 2000ASSL..254...95E}. According to “plate tectonic” proposed by \cite{1991ApJ...366..261R}, the persistent shift  has linear dependence of glitch size found by \cite{Zheng2024}. However our analysis evidently disfavors the linear model explanation of observed data, which means that Bayesian evidence rules out  starquake hypothesis. This convinces us that possible physics inside neutron stars should be taken seriously. Our future work will focus on finding real post-glitch physics of the Crab pulsar by the understandings of these models.
\section{Software and third party data repository citations} \label{sec:cite}

\textit{CompactObject package: An full-scope open-source Bayesian inference framework especially designed for Neutron star physics} \url{https://github.com/ChunHuangPhy/CompactOject}, documentation: \url{https://chunhuangphy.github.io/CompactOject/}.
\textit{UltraNest}: \cite{2021JOSS....6.3001B}, \url{https://github.com/JohannesBuchner/UltraNest}.

\begin{acknowledgments}
This paper is supported  by the National SKA Program of China (Grant No. 2020SKA0120300) and the National Natural Science Foundation of China (Grant Nos. 12033001,12473039). C.H. also acknowledges support from an Arts \& Sciences Fellowship at Washington University in St. Louis and from NASA grant 80NSSC24K1095. 
\end{acknowledgments}

\bibliography{bibfile}{}
\bibliographystyle{aasjournal}



\end{document}